\documentclass[aip,reprint,twocolumn, floatfix, superscriptaddress,dvipdfmx]{revtex4-1}
\usepackage{amsmath,bm,graphicx}
\usepackage{amsmath}
\usepackage{amssymb}
\usepackage{amsfonts}
\usepackage{color}
\begin{document}
\title{Shape dependence of resistance force exerted on an obstacle placed in a gravity-driven granular silo flow}

\author{H. Katsuragi}
\affiliation{Department of Earth and Environmental Sciences, Nagoya University, Nagoya 464-8601, Japan}
\author{K. Anki Reddy}
\affiliation{Departiment of Chemical Engineering, Indian Institute of Technology Guwahati, Guwahati 781039, India}
\author{K. Endo}
\affiliation{Department of Earth and Environmental Sciences, Nagoya University, Nagoya 464-8601, Japan}

\date{\today}

\begin{abstract}
Resistance force exerted on an obstacle in a gravity-driven slow granular silo flow is studied by experiments and numerical simulations. In a two-dimensional granular silo, an obstacle is placed just above the exit. Then, steady discharge flow is made and its flow rate can be controlled by the width of exit and the position of obstacle. During the discharge of particles, flow rate and resistance force exerting on the obstacle are measured. Using the obtained data, a dimensionless number characterizing the force balance in granular flow is defined by the relation between the discharge flow rate and resistance-force decreasing rate. The dimensionless number is independent of flow rate. Rather, we find the weak shape dependence of the dimensionless number. This tendency is a unique feature for the resistance force in granular silo flow. It characterizes the effective flow width interacting with the obstacle in granular silo flow.
\end{abstract}
\maketitle

\section{Introduction}
In granular silo flow, discharge flow rate is a function of exit width and particle size~\cite{Beverloo1961}. As long as the flow is not arrested by the arch formation in the exit region, the discharge flow is stable independent of the thickness of granular layer left in the silo. This particular feature of granular silo flow enables us to produce good hourglasses. However, one of the most serious problems in granular silo flow is clogging. To prevent the clogging, an obstacle placed in front of the exit is sometimes used. The position and shape of obstacle have to be carefully controlled to effectively prevent the clogging~\cite{Zuriguel:2011wc,Endo2016}. In our previous study~\cite{Endo2016}, we have investigated the effect of obstacle shape and its position to the clogging prevention. We have found that triangular or horizontal-bar obstacle is efficient to prevent the clogging by reducing the local packing fraction at the exit region. This implies that the interaction between granular silo flow and obstacle is not very simple. We have to understand their complex relation. Particularly, the resistance force exerted on the obstacle must be a useful quantity to characterize the relation between granular silo flow and obstacle. 

Drag coefficient has long been an important quantity to characterize the interplay between a solid object and fluid flow. In almost all fundamental textbooks on fluid mechanics, the drag coefficient has been explained (e.g.~\cite{Landaufluid,KunduFluid}). When a spherical obstacle of diameter $D$ is placed in a viscous flow of density $\rho_f$ and speed $v$, drag coefficient $C_D$ is defined as $C_D = 2F_D/ \rho_f v^2 A$, where $F_D$ and $A=\pi D^2/4$ are the drag force exerted on the obstacle and cross-sectional area perpendicular to the flow, respectively. This form can readily be extended to various-shape obstacle cases. Indeed, it is well known that $C_D$ strongly depends on the shape of obstacle. The relation between $C_D$ and obstacle shape is a crucial factor in fluid engineering because the smaller $C_D$ in high Reynolds number ($R_e=\rho_f v D/\eta$ , where $\eta$ is viscosity) regime is favorable for energy-saving vehicles design~\cite{HoernerFluid,CengelFluid}. Moreover, physics of resistance force is a key factor not only in engineering but also in astrophysical research. In space, for example, interaction between solid particles and gas flow determines the growth process of dust aggregates in planetesimal formation process~\cite{Armitage2010}. 

However, the relevance of drag coefficient to the resistance force exerted on the obstacle in granlar silo flow has not been studied well so far. Because granular flow is quite different from Newtonian viscous flow, the form of useful dimensionless number characterizing the relation between flow and obstacle could be very different. 
Recently, resistance force created by granular flow has also been extensively studied~\cite{Katsuragi2016}. Granular resistance-force characterization relates to various phenomena such as impact cratering~\cite{Katsuragi:2007fu,Katsuragi:2013ho,Katsuragi2017}, effective swimming~\cite{Shimada2009,Maladen2009}, and bulldozing~\cite{Gravish2010}. In recent literatures, the effects of inertial drag~\cite{Katsuragi:2007fu}, jamming transition~\cite{Takehara2014}, and interstitial fluid~\cite{Royer2011} etc. have been reported in relatively high-speed flow regime. In addition, the effects of hydrostatic pressure~\cite{Albert:1999fs,Stone2004b}, air fluidization~\cite{BrzinskiIII:2010fx}, scaling relation~\cite{Kumar2017}, and memory~\cite{Guillard:2013eu} etc. have also been studied in slow and/or fast-flow regimes. In the slow regime, a weak effect of obstacle shape on granular resistance force has also been reported~\cite{Albert:2001hs}. In this study, we are going to focus on the resistance force acting on the obstacle in granular silo flow. To properly characterize the details of granular resistance force, a dimensionless form should be established. In particular, the dimensionless form would be useful to characterize the obstacle-shape dependence of the resistance force. Such a dimensionless form could also be helpful to compare granular-silo-flow resistance with other soft matter resistance forms.  Therefore, we perform a set of simple experiments and numerical simulations by which a type of dimensionless number relevant to granular flow characterization in slow regime can be defined. Using the dimensionless number, the obstacle-shape-dependent granular flow state in silo can be discussed. 

\section{Methods}
\subsection{Experiment}
\begin{figure}
\begin{center}
\includegraphics[width = 2.75 in]{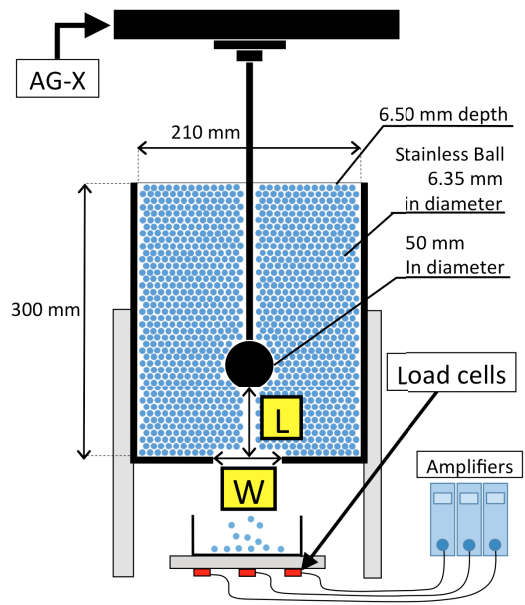}
\end{center}
\caption{Schematic image of the experimental apparatus. A two-dimensional rectangular silo is built with acrylic plates and aluminium bars. A solid obstacle (sphere ($50$~mm diameter), triangle ($50$~mm side), or inverted triangle ($50$~mm side)) is hung by a universal testing machine. After the silo is filled with stainless-steel spheres (6.35~mm diameter), an exit of width $W$ is opened at the center of bottom wall. During the discharge, resistance force exerted on the obstacle $F$ and discharge flow rate $Q$ are measured. $W$ and the distance from exit to obstacle $L$ are the main control parameters in this experiment.}
\label{fig1}
\end{figure}

\begin{figure*}
\includegraphics[scale=1.]{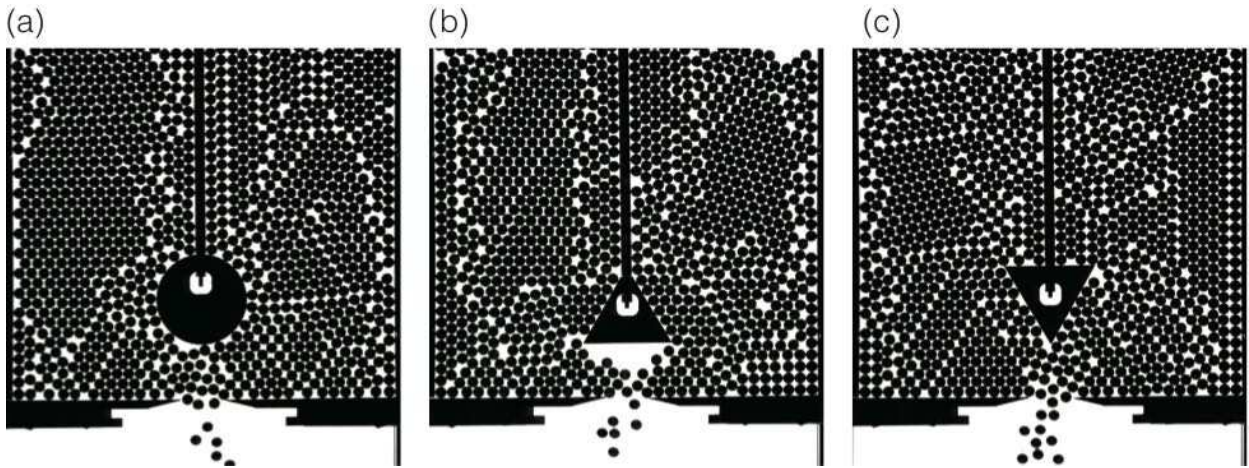}
\caption{Example snapshots of grain configurations with (a)~circular, (b)~triangular, and (c)~inverted-triangle obstacles. Parameters are $W=25$~mm and $L=30$~mm. Although some partially-ordered structures can be observed since we use monodisperse particles, the structure is not perfectly ordered. }
\label{fig:raw_imgs}
\end{figure*}

Experimental apparatus is schematically shown in Fig.~\ref{fig1}. The system is basically identical to that used in~\cite{Endo2017,Endo2016}. We built a two-dimensional (2D) granular silo with a horizontal bottom wall. Two acrylic plates sandwich aluminum bars of thickness $6.5$~mm (inner dimension of the cell: $300\times210$~mm). An obstacle is hung by a universal testing machine (Shimadzu AG-X) via a stainless-steel rod in diameter $6$~mm. This stainless rod partitions the silo room and causes sidewall effect to the flow. Although this rod could affect the flow behavior, we consider its effect must be minor because the result by numerical simulation, in which the rod is absent, agrees well with the experimental result, as discussed later. To observe the shape effect, we employ three kinds of obstacles: circle (diameter $50$~mm, thickness $6$~mm), triangle (length of one side $50$~mm, thickness $6$~mm) and inverted triangle (same size as triangle). This experimental cell is filled with stainless-steel spherical particles of diameter $d=6.35$~mm and density $\rho=7950$~kg/m$^3$. In order to minimize the particle-wall friction, spherical particles are used in the 2D experiment. After the filling, an exit is opened at the center of bottom wall. Then, the discharge flow is made as a silo flow. The flow rate can be controlled by both the width of exit $W$ and the vertical distance between the exit and bottom of obstacle $L$. When $W$ or $L$ is too small, clogging often occurs~\cite{Janda:2008bq,Zuriguel:2011wc}. In this study, however, we focus on the steady-flow regime to characterize the granular resistance force exerted on the obstacle. Thus, only the steady flow duration is analyzed; data after clogging are discarded. We confirm that the discharge flows are always steady even just before the clogging. Actual examples of particle configurations with circular, triangular, and inverted-triangle obstacles are shown in Fig.~\ref{fig:raw_imgs}. Since monodisperse particles are used in this study, partially ordered structures can be observed. However, the typical flow features for granular silo flow (e.g., steadiness independent of layer's thickness) are preserved in this system. More detail characterization and the study on clogging with this experimental system are reported in~\cite{Endo2016,Endo2017}. The videos of particle motions can be found in~\cite{Endo_video}. The variation of $W$ ranges from $25$ to $60$~mm. In each $W$, $L$ is varied at most from $100$ to $0$~mm. The actual lower limit of $L$ depends on the shape of obstacle. Discharge mass flow rate $Q$ and resistance force exerted on the obstacle are acquired by load cell sensors (Kyowa LMB-A) placed beneath the dish collecting particles and the load cell attached to the testing machine, respectively. At least three experimental runs are performed in each experimental condition. Note that, in this experiment, particles are never replenished during the discharge. Namely, the number of particles within the silo monotonically decreases during the discharge. Nevertheless, the flow rate is almost constant just like hourglasses. By simple analysis of the obtained data, we will define a dimensionless number characterizing the granular slow-flow state. Using the dimensionless number, the obstacle-shape dependence of resistance force will also be discussed. 

\subsection{\label{sec:simu_method} Numerical simulation}
 We use Discrete Element Method (DEM) \cite{cundall79,brilliantov96,silbert2001} to simulate the gravity-driven granular flow discussed in the present work. In this method, classical equations of motion (linear and angular momentum balances) will be integrated for each particle to follow its motion in time. This requires us to know how much force or torque a particle  experiences because of its interaction with other particles or walls present in the system. We briefly explain here  the computation of these forces and torques. Let us consider two spherical particles \textit{i} and  \textit{j} in contact and $\textbf{n}$ is a unit vector along the line joining the centers of the particles. The relative velocity at the contact point can be given as 
$\textbf{v}_{ij} = \textbf{v}_{i} - \textbf{v}_{j} + (R_{i} \omega_{i} + R_{j} \omega_{j})  \times \textbf{n}$. Here $\textbf{v}_{i}$, $R_{i}$ and $\omega_{i}$ are the linear velocity, radius, and angular velocity of particle \textit{i}.
Hence relative velocity in the normal direction is $\textbf{v}_{ijn}=(\textbf{v}_{ij}\cdot\textbf{n})\textbf{n}$  and tangential direction is $\textbf{v}_{ijt} = \textbf{v}_{ij} - \textbf{v}_{ijn}$. As DEM uses spring and dashpot models, forces depend on the overlap and the relative velocity between particles \textit{i} and \textit{j}. For a Hertzian contact model, force in the normal direction is given by $\textbf{F}_{n} = \sqrt{R_{\textrm{eff}}\delta}(k_{n}\delta \textbf{n}-m_{\textrm{eff}}\gamma_{n}\textbf{v}_{ijn})$, where $k_{n}$, $\gamma_{n}$ and $\delta$ are the stiffness coefficient (spring constant), damping coefficient, and overlap in the normal direction, respectively. Overlap is given by $R_{i} + R_{j} - (\textbf{r}_{j}-\textbf{r}_{i})\cdot\textbf{n}$, where $\textbf{r}_{i}$ and $\textbf{r}_{j}$ are the position vectors of particles \textit{i} and \textit{j}. $R_{\textrm{eff}}$ is defined as $\frac{R_{i} R_{j}}{(R_{i}+ R_{j})}$ and $m_{\textrm{eff}}$ is defined as $\frac{m_{i} m_{j}}{m_{i}+m_{j}}$, where $m_{i}$ and $m_{j}$ are the masses of particles \textit{i} and \textit{j}. Force in the tangential direction is given by $\textbf{F}_t=-\textrm{min}(\mu \textbf{F}_n,\sqrt{R_{\textrm{eff}}\delta} (k_t\Delta {s_t}+m_{\textrm{eff}}\gamma_{t} \textbf{v}_{ijt}))$. Here $\mu$, $k_{t}$, $\Delta s_{t}$, and $\gamma_{t}$ are the coefficient of friction, elastic constant for tangential contact, tangential displacement vector between particles \textit{i} and \textit{j}, and viscoelastic damping constant for tangential contact, respectively. Torque exerted on the particle \textit{i} can be computed as $R_{i} \textbf{n} \times \textbf{F}_{t}$. In the present simulation, all the lengths are scaled by the particle diameter $d$. Time, force, and stress are normalized to $\sqrt{d/g}$, $\rho d^{3} g$, and $\rho d g$, where $\rho$ and $g$ are density of the particle and acceleration due to gravity. Both $\rho$ and $g$ are taken as unity in the simulation. Spring and damping coefficients ($k_{n}=2\times 10^{8} \rho d g, k_{t}=2.25\times10^{8} \rho d g, \gamma_{n}=\gamma_{t}=10000 \sqrt g/d^{1.5}$) used as parameters in the present force model are taken to represent the materials used in the experimental study. To compare the numerical result with experiments, we have to substitute the specific values: $\rho=7950$~kg/m$^3$, $g=9.8$~m/s$^{-2}$, and $d=6.35$~mm. Coefficient of friction $\mu$ has been set as 0.36 for particle-particle interaction, while it is kept as 0.5 for particle-wall interactions.

\begin{figure}
\includegraphics[scale=0.45]{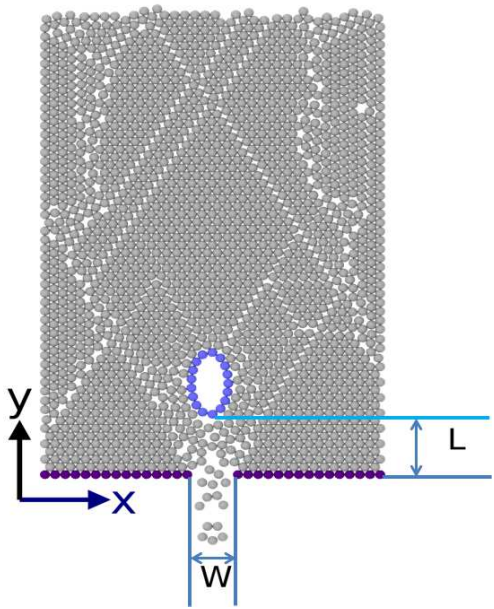}
\caption{\label{fig:simusnap} Simulation system with an ellipse obstacle (blue color) and particles. Here width of outlet ($W$) is $4 d$, distance between the bottom of obstacle and the exit ($L$) is $6.3 d$.}
\end{figure}

To simulate the granular flow past obstacles, first we fix the obstacle at a relevant position and pour the particles into a container under the gravity and we ensure that kinetic energy of total system is very close to zero prior to discharge of granular particles. There are 1900 particles in the silo  which is in correspondence with the number of  particles used in the experiment. This results in a simulation box of $34d \times 48d \times 1d$. The simulation system is shown in Fig.~\ref{fig:simusnap}. Flat frictional walls are there in the x-direction (right and left). We allow the particles to discharge through the opening of a silo with a certain width $W$. The bottom wall consists of the fixed particles. As the particles flow past stationary obstacle, the force experienced by the obstacle is recorded at regular intervals. 
Similarly we also record the number of particles discharged to compute the flow rate $Q$ like in the experiment. Particles are not replenished also in the numerical simulation. 
All the simulations are carried out using the Large Atomic Molecular Massively Parallel Simulator (LAMMPS) \cite{plimpton95} and the visualizations are done using Visual Molecular Dynamics (VMD) \cite{humphrey96}. 
The monodisperse system is employed in numerical simulation since the experiment also uses approximately monodisperse particles. We use a circle, a triangle, an inverted triangle, horizontal bars, and ellipses as the obstacles. These shapes consist of fixed particles whose size is identical to the flowing particles~\cite{Endo2016}. Except the ellipses and horizontal bars, the horizontal length (width) of the obstacle is always set $8d$ which is close to the width of obstacles used in the experiment, $50$~mm. Two aspect ratios ($1.584$, $3.146$) with a fixed width $4.62d$ are employed for the study of flow past ellipse shape. Here, the longer axis of the ellipse is aligned to the vertical direction as shown in Fig.~\ref{fig:simusnap}. Various lengths ($6d$, $8d$, and $10d$) of horizontal bars are used to observe the effect of horizontal dimension of the obstacle.

\section{Results and analyses}
\subsection{Experimental results and analyses}
\begin{figure}
\begin{center}
\includegraphics[width = 3.37 in]{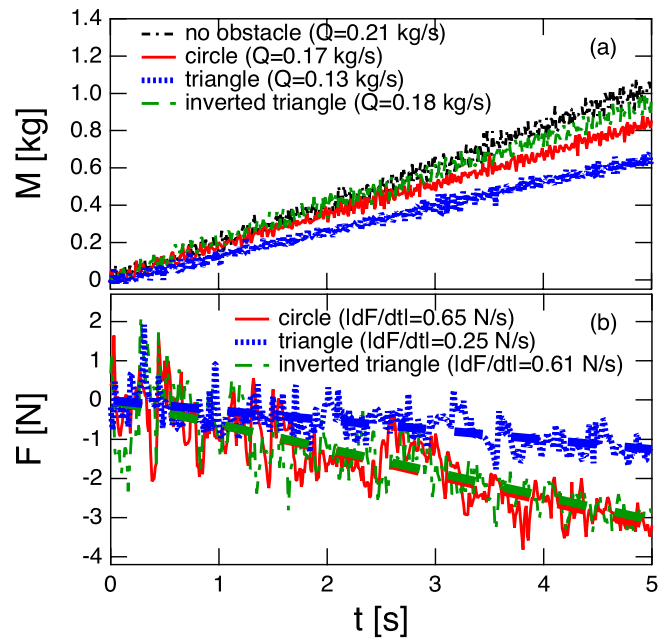}
\end{center}
\caption{Experimentally obtained raw time-series data of (a) discharged mass $M$ and (b) resistance force difference $F$. The slopes in the plots (a) and (b) correspond to $Q$ and $dF/dt$, respectively. Color and line codes indicate the obstacle shape as denoted in the legend. $W$ and $L$ are fixed to $25$~mm and $30$~mm, respectively. }
\label{fig2}
\end{figure}

First, we show the flow rate data in Fig.~\ref{fig2}(a). The data shown in Fig.~\ref{fig2} are taken with the fixed experimental conditions: $W=25$~mm and $L=30$~mm (or no-obstacle). Discharged mass $M$ with various obstacles as well as no-obstacle case is plotted as a function of time $t$. One can clearly confirm the nice linearity for all the flow rates shown in Fig.~\ref{fig2}(a). The values of correlation coefficient are greater than $0.99$ for all relations. This indicates that the discharge flow is steady even under the influence of obstacle. This steady flow rate can actually be observed under all experimental conditions. The flow rate $Q$ is measured by the slope in this plot. As can be seen in Fig.~\ref{fig2}(a), the existence of obstacle decreases the flow rate. The degree of decrease depends on the shape of obstacles. Triangular obstacle decreases the flow rate most significantly. This small flow rate with a triangular obstacle can be related to the effective clog prevention. Detail discussion on the clog prevention by a triangular obstacle can be found in our previous paper~\cite{Endo2016}. Here, in this paper, we are going to discuss the resistance force exerted on the obstacle in granular silo flow.

The corresponding resistance force as a function of time $t$ is shown in Fig.~\ref{fig2}(b). In this plot, $F$ indicates the resistance-force difference from the initial state. Since the initial value of the measured force significantly fluctuates depending on packing protocols, here we only focus on the flowing state to discuss the resistance-force behavior in steady flow regime rather than the absolute force behavior including its initial state. Thus, we analyze the resistance-force difference $F(t)=f(t)-f(0)$, where $f(t)$ and $f(0)$ are the instantaneous resistance force and its initial value, respectively. Perhaps, the large variance of the initial force might affect the behavior of resistance-force difference $F(t)$ in the flowing state. However, we confirm the reproducibility of $F(t)$ behavior with the same experimental conditions but different initial configurations. Actually, the characterization of steady flowing regime is usually easier than that of static state in various granular behaviors (see e.g.~\cite{Bertho:2003jx,Katsuragi2015,Katsuragi2016}). The current result is a typical example of such complex granular behaviors. In Fig.~\ref{fig2}(b), $F(t)$ shows relatively large fluctuation compared to $M(t)$ shown in Fig.~\ref{fig2}(a). The values of correlation coefficient for these data are approximately $-0.7$. This fluctuation comes from the series of discrete collisions among flowing particles and obstacle. In this study, however, we would like to focus rather on the average behavior of resistance force in the steady granular silo flow. 
The linear decreasing of $F(t)$ implies that the hydrostatic pressure plays more crucial role for the resistance force than rate-dependent dynamic pressure of granular flow. This is a little counterintuitive because the flow rate $Q$ is independent of the hydrostatic pressure in granular silo flow. 
Even in relatively large $Q$ regime, it is difficult to observe the clear rate dependence of the resistance force in this experiment. It always relates to the force decreasing rate. To characterize such a flow state, here we introduce a simple dimensionless number. 

\begin{figure}
\begin{center}
\includegraphics{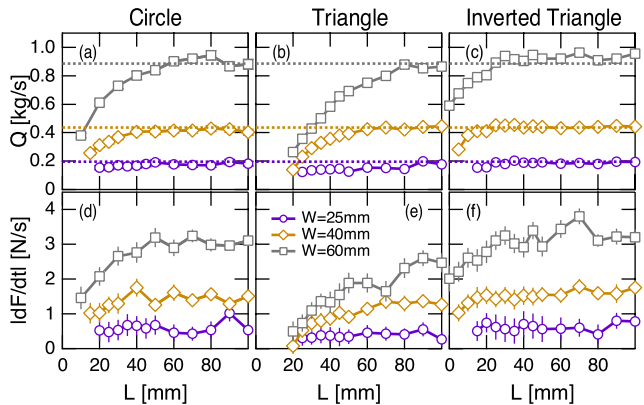}
\end{center}
\caption{Experimentally obtained flow rates $Q$ and resistance-force decreasing rates $|dF/dt|$. Upper and bottom rows indicate $Q$ and $|dF/dt|$, respectively. Left, center, and right columns correspond to circular, triangular, and inverted-triangle obstacle cases, respectively. Colors and symbols represent $W$ values. Horizontal dotted lines in the upper-row plots denote the level of $Q$ without obstacle, $Q_{\rm no}$. While the data of $W=30$~mm are not shown, they show qualitatively similar trend. Error bars indicate standard errors of three experimental runs.}
\label{fig3}
\end{figure}

To compute the dimensionless number, $Q$ and $|dF/dt|$ are systematically measured by varying $W$ and $L$. The measured results are shown in Fig.~\ref{fig3}. The flow rates $Q$ for (a) circle, (b) triangle and (c) inverted triangle are shown in the upper row. And the force decreasing rates $|dF/dt|$ for (d) circle, (e) triangle, and (f) inverted triangle are presented in the bottom row. All the data are shown versus $L$, and colors and symbols indicate $W$ values as denoted in the legend. Horizontal dotted lines indicate the flow-rate levels in the no-obstacle case, $Q_{\rm no}$. Obviously, $Q$ approaches $Q_{\rm no}$ when $L$ is sufficiently large. This result is natural because $L=\infty$ corresponds to the no-obstacle case. In large $W$($=60$~mm) cases, the characteristic distance $L_c$, at which $Q$ becomes almost identical to $Q_{\rm no}$, strongly depends on the obstacle shape; $L_c$ is approximately $60$, $80$ and $30$~mm for circle, triangle, and inverted triangle, respectively. Basically, $Q$ is an increasing function of both $W$ and $L$. One can confirm a slight peak of $Q$ at $W=60$~mm and $L=80$~mm in the circular obstacle case~(Fig.~\ref{fig3}(a)). Although this behavior is similar to the peak of flow rate observed in previous study~\cite{Lozano2012} in which the maximum (peak) of $Q$ is observed by varying $L$, the peak trend in Fig.~\ref{fig3}(a) is not very clear. The resistance-force decreasing rate $|dF/dt|$ shows qualitatively similar behaviors as shown in Fig.~\ref{fig3}(d-f). This is reasonable since the larger $Q$ results in the faster discharge. In other words, the number of particles above the obstacle rapidly decreases when $Q$ is large. If only the fluid-like isotropic hydrostatic pressure dominates $F$ even in granular silo flow, the relation between $|dF/dt|$ and $Q$ should be universal. 

\begin{figure}
\begin{center}
\includegraphics[width = 3.37 in]{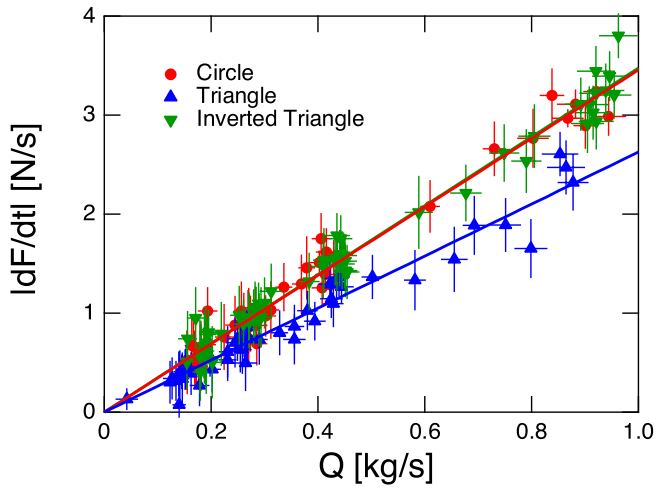}
\end{center}
\caption{Relation between $|dF/dt|$ and $Q$ in the experiment. All the experimental data are displayed in this plot. $|dF/dt|$ is simply proportional to $Q$. And the factor of proportionality depends on the shape of obstacle. Error bars are same as those in Fig.~\ref{fig3}.}
\label{fig4}
\end{figure}

To check the shape dependence of the relation between $|dF/dt|$ and $Q$, we directly compare them in Fig.~\ref{fig4}. As shown in Fig.~\ref{fig4}, they indeed have proportional relations. However, the factor of proportionality depends on the shape of obstacle. Actually, the linear relation itself is not very surprising. From the dimensional thought, $|dF/dt|$ and $Q=|dM/dt|$ should obey the relation,
\begin{equation}
\left| \frac{dF}{dt} \right| = gC_{SD}Q.
\label{eq:C_SD}
\end{equation}
In this form, the dimensionless number $C_{SD}$ characterizes the relation between resistance force and flow state. Note that $C_{SD}$ can only be applied to the resistance force acting to an obstacle in a granular silo flow. The form is quite different from the conventional dimensionless number characterizing the resistance force such as drag coefficient. The advantage of usage of the silo system to discuss the resistance-force behavior is its easiness of the control of steady flow rate. While a certain complexity of granular flow might affect the relation between $|dF/dt|$ and $Q$, the linear relation can be confirmed in all the obstacle-shape cases. By the least square fitting, $C_{SD}$ for circle, triangle, and inverted triangle are computed as $0.353 \pm 0.005$, $0.268 \pm 0.005$, and $0.355 \pm 0.003$, respectively. The triangle's $C_{SD}$ is less than other two. However, the variation of $C_{SD}$ is not very significant. This slight difference could be a key to understand the obstacle-shape-dependent granular silo flow field.

If $Q$ is too large, the value of $F$ should directly depend on $Q$. In such a relatively fast-flow regime, the resistance force will be a quadratic function of flow speed; inertial drag regime~($F_D\sim \rho v^2 A$). In the current experimental conditions (slow granular silo flow), however, the flow rate rather relates to the decreasing rate of resistance force. And the relation depends on the shape of obstacle. 

\subsection{Numerical results and analyses}
\begin{figure*}
\includegraphics[scale=1.2]{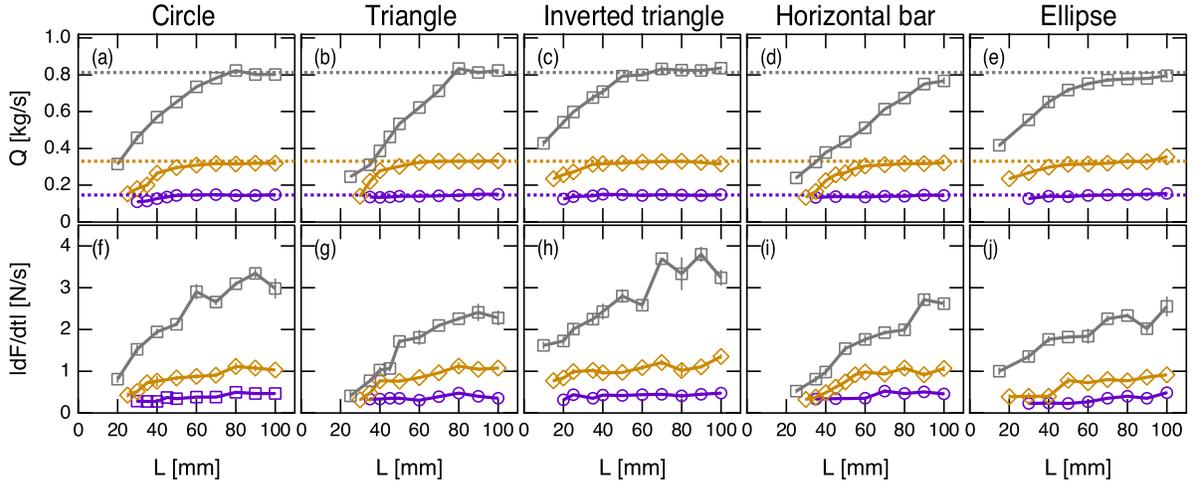}
\caption{\label{fig:flow-shape} (a-e) Flow rates $Q$ and (f-j) $|dF/dt|$ values from the numerical study. (Gray $(W=10d)$, yellow $(W=6d)$ and purple $(W=4d)$.) These widths correspond to $W=60$~mm, $W=40$~mm, and $W=25$~mm in the experimental study. Plots of ellipse (e,j) correspond to the smaller ellipse (aspect ratio 1.58) case. Horizontal lines in the plots of top row represent the flow rates observed in a silo without obstacle, $Q_{\rm no}$. }
\end{figure*}

Here we try to reproduce the above experimental observations by numerical simulations. In the present numerical study, three widths of the silo are considered ($W=4d$, $6d$ and $10d$)  and the distance between the bottom of obstacle and exit varied  from $L=1.6 d$ to $L=15.8 d$ in correspondance with the experimental study. $Q$ and $|dF/dt|$ for various obstacle shapes are shown  in Fig.~\ref{fig:flow-shape}, which are in good agreement with  experimental data. Here, to directly comare the values, we multiply $(\pi/6)\rho d^3 (d/g)^{-1/2}$, $(\pi/6)\rho d^3 g (d/g)^{-1/2}$, and $d$ to $Q$, $|dF/dt|$, and $L$, respectively. When the obstacle is sufficiently far from the exit, there is no effect on the flow rate; $Q\simeq Q_{\rm no}$. $Q_{\rm no}$ values are represented by horizontal dashed lines in Fig.~\ref{fig:flow-shape}. As $L$ approaches near the exit, there is a significant reduction in the flow rate for all the widths of exit considered here. The behavior of $|dF/dt|$ is similar to that of $Q$. That is, all the numerical results successfully reproduce the experimental results. 

\begin{figure}
\includegraphics[width = 3.37 in]{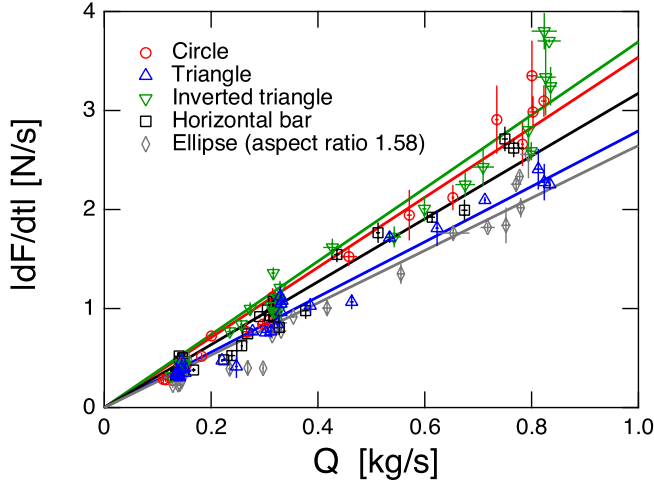}
\caption{\label{fig:cd-all} $|dF/dt|$ versus $Q$ from the numerical simulation for the granular flow past obstacles of various shapes. Here the lines represent fits to Eq.~(\ref{eq:C_SD}).}
\end{figure}

The values of $C_{SD}$ for all the shapes are obtained with the help of fits to $|dF/dt|$ versus $Q$ by Eq.~(\ref{eq:C_SD}) as shown in Fig.~\ref{fig:cd-all}. The values of $C_{SD}$ from the simulation study were found to be $0.361\pm 0.0097$ for circle, $0.377 \pm 0.014$ for inverted triangle, $0.285 \pm 0.0093$ for triangle, $0.286 \pm 0.014$ for short ($6d$) horizontal bar, $0.324 \pm 0.012$ for medium ($8d$) horizontal bar, and $0.373 \pm 0.011$ for long ($10d$) horizontal bar. In Fig.~\ref{fig:cd-all}, only the $8d$-width data are shown for the horizontal bar. For ellipse of aspect ratio $1.584$, $C_{SD}$ is $0.270 \pm 0.011$ while it is $0.302 \pm 0.010$ for ellipse of aspect ratio $3.146$ (not shown in the plot). The values of $C_{SD}$ obtained in this study are listed in Table~\ref{tab:C_SD}.
 
\begin{table}
\begin{tabular}{lcc} \hline
Shape & Experiment & Simulation \\ \hline\hline
Circle & 0.353 & 0.361  \\
Triangle & 0.268 & 0.285  \\
Inverted triangle & 0.355 & 0.377  \\
Ellipse (short) & - & 0.270  \\
Ellopse (long) & - & 0.302  \\
Bar (short $6d$ width) & - & 0.286 \\
Bar (medium $8d$ width) & - & 0.324  \\
Bar (long $10d$ width) & - & 0.373 \\ \hline
\end{tabular}
\caption{Values of $C_{SD}$ obtained by experiment and numerical simulation}
\label{tab:C_SD}
\end{table}

\section{Discussion}
One of the most striking features of the dimensionless number $C_{SD}$ is the weak obstacle-shape dependence. Here we briefly discuss the difference between $C_{SD}$ and conventional dimensionless number characterizing flow resistance: drag coefficient. 
$C_{SD}$ varies in the range from 0.27 to 0.37, which is not a significant variation when compared with variation in drag coefficient associated with the viscous flow past obstacles. For a viscous fluid flow past a streamlined obstacle, drag  coefficient will be less compared to the blunt obstacles of same cross section for the reason such as the delay in boundary layer separation or the reduction in the area of low pressure region behind the obstacle~\cite{katz2010}. In the present study also, we observe that ellipse-shaped obstacle has slightly lower value of $C_{SD}$ compared to other shapes of obstacles. However, that is not significant reduction. To access the dynamical regime, much faster flow has to be collided to the obstacle. Then, the shape dependence of the resistance force could be enhanced. In particular, the effect of the tip should be enhanced in relatively fast flow resistance~\cite{Clark2014}. In the current experiment, the central rod hanging the obstacle effectively erases the effect of tip. In addition, the tip singularity is blurred even in numerical simulation since we build the obstacle by particles. Much faster silo flow against the true tip structure is an important future topic to be studied. The current experiment and simulation still correspond to the slow-flow regime.

The current experimental and numerical results are consistent with horizontal slow resistance measurement performed by Albert et al.~\cite{Albert:2001hs} in which the granular slow resistance force weakly depends on the obstacle shape. Furthermore, Albert et al. observed that the resistance force depends on the length (parallel to the flow direction) of obstacle. According to their result, the longer obstacle results in the larger resistance force. This result is qualitatively consistent with our numerical simulation. The longer ellipse causes the larger $C_{SD}$ in our simulation. However, the difference is not very significant again. More systematic study is necessary to reveal the further details of granular resistance force under the influence of obstacle.

Although its shape dependence is limited, physical meaning of $C_{SD}$ can be discussed from its shape-dependent variation. From the viewpoint of effective flow width, the ratio between {\it loading-band width} affecting the resistance force and the entire flow width should correspond to $C_{SD}$. If all the mass of discharging particles contributes to the resistance-force decreasing, $C_{SD}$ should be unity. The smaller $C_{SD}$ implies the fewer particles relate to the resistance-force decrease. Namely, $C_{SD}$ represents the ratio, 
\begin{equation}
C_{SD}=\frac{W_{\rm band}}{W_{\rm flow}}, 
\label{eq:C_SD_W}
\end{equation}
where $W_{\rm band}$ and $W_{\rm flow}$ are the loading-band width and entire flow width, respectively. At present, we do not know the specific value of $W_{\rm flow}$. This must be related to the flow rate $Q$ and the system size. Although the current system size is supposed to be large enough to neglect the sidewall effect in granular silo flow~\cite{Hirshfeld:2001wf}, the system-size dependence of $C_{SD}$ would be very important next step to further characterize the interaction between obstacle and granular flow. In addition, the size ratio between the obstacle and particles might affect the result just like granular frictional resistance~\cite{Furuta2017}. The structural ordering could also affect the behavior of $C_{SD}$ since this study uses monodisperse particles. Systematic series of experiments with various particle sizes should be performed to reveal the universality of $C_{SD}$ behavior. If we assume that $W_{\rm flow}$ can be approximated by the sytem width $210$~mm, the effective loading-band width $W_{\rm band}$ approximately distributes from $56$ to $75$~mm depending on the obstacle shape. These values are comparable to the characteristic length scale (width) of obstacles, $50$~mm. That is, mass of particles within this loading-band width effectively burdens the obstacle during the discharge. By comparing $C_{SD}$ obtained by numerical simulations using horizontal bars of different widths~(Table~\ref{tab:C_SD}), we can confirm that $C_{SD}$ has a positive correlation with the horizontal width. This tendency is qualitatively consistent with the discussion so far. However, $C_{SD}$ could not be simply proportional to the horizontal width, particularly in the small width regime. This implies that $W_{\rm band}$ (and/or $W_{\rm flow}$) nonlinearly depends on the obstacle dimension. To reveal the detail of nonlinear relation among obstacle dimension, $C_{SD}$, $W_{\rm band}$, and $W_{\rm flow}$, systematic measurements with a much larger silo are necessary.

It is well known that the force network structure in granular matter scatters the vertical loading to horizontal direction. Although the current system is shallow (or wide) enough to neglect this so-called Janssen effect~\cite{Janssen:t4ITFCJT}, this force scattering effect could be detected in $C_{SD}$. 
The wider loading-band width means the stronger force scattering. Put differently, the shape dependence of the force network structure in granular flow can also be evaluated using $C_{SD}$ as a loading-band width difference. For circle and inverted triangle, $C_{SD}$ values are almost identical. The value of $C_{SD}$ for triangle is smaller than these two. This tendency is a little counterintuitive because the collision angle between the obstacle surface and vertical flow are different among these three. Furthermore, rather the circle and triangle seem to be similar in terms of the collision angle. Nevertheless, $C_{SD}$ of circle and inverted triangle are quite similar. 

By the oblique collision between granular flow and obstacle, the vertical component of force is effectively scattered to the horizontal direction. Since all the shapes of obstacles used in this study are symmetric, this horizontal scattering of the force effectively compresses the granular-flow column mainly in horizontal direction. This effective compression could cause the slowing of the flow itself by narrowing free space and enhancing dissipation by collisions or friction. In fact, the net flow rate $Q$ for triangular obstacle is slower than circular case~(Fig.~\ref{fig1}). 

The microscopic details of particle motions have not been discussed in this paper. We have only focussed on the macroscopically average quantities: flow rate and resistance force. As seen in Fig.~\ref{fig2}, $F$ shows considerable fluctuation. To characterize the further details of obstacle effect in granular silo flow, fluctuation analysis is an interesing next step. Microscopic characterization to understand the details of this newly defined quantity $C_{SD}$ is an important future problem. For instance, the force chain characterization like Ref.~\cite{Tang:2011hc,Iikawa:2016eg,Iikawa2017} might be a possible way to probe the detail structure of force network in the granular silo flow. Since the simple methodology is developed in this study, consecutive and systematic measurements with various obstacle shapes might provide useful information about the efficient (energy-saving) motion within granular matter. 

\section{Conclusion}
In summary, a dimensionless number characterizing the interplay between granular flow and obstacle has been defined and measured in a gravity-driven granular silo flow. Since the dimensionless number is defined by the ratio between resistance-force decreasing rate and discharge flow state, it can characterize the interaction between granular sio flow and obstacle. Using this dimensionless number, the slightly shape-dependent flow-band width in a slowly flowing dense granular silo can be discussed. The flow-band width could be related to the granular force scattering due to the random network of particles. According to the experimental and numerical results, circle and inverted triangle exhibit almost the same $C_{SD}$ while the triangle and ellipse result in smaller $C_{SD}$. Perhaps, this characterization method  could be a useful evaluation way for designing the efficient shape to move within granular matter. 

\section*{Acknowledgement}
This work was supported by JSPS KAKENHI Grants No.~15H03707 and No.~18H03679. K. Anki Reddy would like to thank IITG start-up research grant.

\bibliography{sd}

\end{document}